\documentclass[a4paper]{jpconf}
\usepackage{graphicx}
\usepackage{epstopdf}

\begin{document}
\title{\bf{(2+1) dimensional cosmological models in $f(R,T)$ gravity with $\Lambda(R,T)$}}

\author
{Safiqul Islam $^a$ , Praveen Kumar $^b$, G.S. Khadekar $^c$\\ and Tapas K Das $^d$\\
	$^{a, d}$Harish-Chandra research Institute, HBNI, Chhatnag Road, Jhunsi, Allahabad-211019, India \\
	$^{b, c}$Department of Mathematics, Rashtrasant Tukadoji Maharaj Nagpur University, Nagpur 440033, India\\
	$^a$safiqulislam@hri.res.in \\ $^b$pkumar6743@gmail.com\\ $^c$gkhadekar@yahoo.com\\
	$^d$tapas@hri.res.in
}


\begin{abstract}
We intend to study a new class of cosmological models in $f(R,T)$ modified theories of gravity, hence define the cosmological constant $\Lambda$ as a function of the trace of the stress energy-momentum-tensor $T$ and the Ricci scalar $R$, and name such a model “$\Lambda(R,T)$ gravity”
where we have specified a certain form of $\Lambda(R,T)$. $\Lambda(R,T)$ is also defined in the perfect fluid and dust case. Some physical and geometric properties of the model are also discussed. The pressure, density and energy conditions are studied both when $\Lambda$ is a positive constant and when $\Lambda=\Lambda(t)$, i.e a function of cosmological time, t. We study behavior of some cosmological quantities such as Hubble and deceleration parameters. The model is innovative in the sense that it has been described in terms of both $R$ and $T$ and display better understanding of the cosmological observations. 
\end{abstract}

\section{Introduction}
~~~It is known that General Relativity is the standard theory of gravity. However alternative equations for the gravitational field are resorted to, when the spacetime dimension is reduced to $2+1$, due to the difficulty in defining a proper Newtonian limit \cite{i}. The $2+1$ dimensional analogue is generated by any circularly symmetric matter distribution. Circularly symmetric distributions of charged matter in $2+1$ dimensions are detailed in \cite{m}, where the authors have noticed that hydroelectrostatic equilibrium for a charged fluid with equation of state $p=p(\rho)$ with $p>0$ takes place only in anti de-Sitter spaces. The cosmological solutions of the $3+1$ dimensional Einstein equations are rather cumbersome and dominated by non-integrability, in contrast the theory in $2+1$ offers the possibility of finding the general solution \cite{n}. A new class of exact interior solutions in $2+1$-dimensional spacetime has been shown in \cite{o}, for the perfect fluid model both with and without the cosmological constant $\Lambda$.

To describe the early universe, the f(R,T) theory of gravity is considered as the fundamental gravitational theory. It has arouse much attention in the last decade and is in fact the generalization of f(R) and f(T) gravitational theories. Various authors have constructed several aspects for both f(R) and f(T) modified theories. The exact vacuum solutions of Bianchi type I, III and Kantowski-Sachs spacetimes in the metric version of f (R) gravity has been studied considering the expansion scalar $\theta$ to be proportional to the shear scalar $\sigma$ \cite{j}. Bengochea and Ferraro proposed to replace the TEGR that is the torsion scalar Lagrangian $T$ with a function $f(T )$ of the torsion scalar, and studied its cosmological consequences \cite{p}. This type of modified gravity is nowadays called as $f(T)$ gravity theory.

The authors in \cite{l} have studied two particular models of f $(R, T )$ gravity namely, $f(R)+\lambda T$ and $R + 2f(T)$ and derived their power-law solutions in homogeneous and isotropic $f(R,T)$ cosmology.

The cosmological constant was introduced by Einstein who later rejected it after expansion nature of the universe was discovered by Hubble. The results of type Ia supernova [\cite{g} ,\cite{h}] show that the universe is accelerating rather than decelerating. These results suggest that our universe can have a non-zero cosmological constant. In GR however it is the dark energy which behaves like a cosmological constant at early time and supports the accelerated expansion of the universe. We can remove this consideration of dark energy in our modified $f(R,T)$ gravity model.

It is observed that the violation of energy conditions in modified gravity indicates the attractive nature of gravity whereas repulsive gravity may occur for ordinary matter that satisfies all the energy conditions \cite{f}.

The organization of this paper is envisaged as follows:
In section II. we generate new solutions under $f(R,T)$ gravity and deduce the cosmological constant as a function of the Ricci scalar $R$, the trace of the energy momentum tensor $T$ and another constant $\lambda$. The field equations and solutions of the generalized metric in (2+1)-Dimensional Spacetime are shown in section III. Here we study the energy conditions and various consequences when $\Lambda$ is treated both as a constant and as a function of cosmic time t, i.e $\Lambda=\Lambda(t)$. We find the scale factor, the Hubble and deceleration parameters and observe their changes with time, t. The study ends with a concluding remark.

\section{Generating new solutions under $f(R,T)$ models:}

We consider the modified gravity action as,
\begin{eqnarray}
	S=\frac{1}{16 \pi} \int (\sqrt{-g}f(R,T)+ \sqrt{-g}\mathcal{L}_m) d^3 x,
\end{eqnarray}
where f(R,T) is an arbitrary function of the Ricci scalar, R, T is the trace of the stress-energy momentum tensor, $T_{\mu \nu}$ given by $T=g^{\mu \nu} T_{\mu \nu}$ and $\mathcal{L}_m$ stands for the matter Lagrangian density. $T_{\mu \nu}$ is defined as,
\begin{eqnarray}
	T_{\mu \nu}=-\frac{2}{\sqrt{-g}} \frac{\delta \sqrt{-g} \mathcal{L}_m}{\delta g^{\mu \nu}},
\end{eqnarray}

Several models of f(R,T) gravity have been studied by various authors, depending on the matter source. They may be illustrated as follows,
\begin{eqnarray}
	(i) f(R,T)= R+ 2f(T)~~~~~~~~~~~~~~~~\nonumber\\
	(ii) f(R,T)= f_{1}(R)+ f_{2}(T)~~~~~~~~~~~\nonumber\\
	(iii) f(R,T)= f_{1}(R)+ f_{2}(R) f_{3}(T)~~~\nonumber\\
	(iv) f(R,T^{\phi})~~~~~~~~~~~~~~~~~~~~~~~~~~~~~~~
\end{eqnarray}
where $\phi$ is a scalar field.

The cosmological intricacy in (i) where $f(R,T)= R+ 2f(T)$ has been discussed by authors \cite{c}, \cite{d}. The cosmological model in (ii) is developed in \cite{k} to effectively discuss the transition from the matter dominated phase to an accelerated phase. With the model as (ii ), a new class of solutions pertaining to some cosmic scale functions have been investigated in \cite{b}, where the authors study the cosmological constant $\Lambda$ as a function of $T$.

We consider the f(R,T) model given by (iii). The gravitational field equations with perfect fluid matter source are given by \cite{a},
\begin{eqnarray}
	[f_1'(R)+f_2'(R) f_3(T)]R_{\mu \nu}-\frac{1}{2}f_1(R)g_{\mu \nu}
	+(g_{\mu \nu}\makebox-\nabla_{\mu}\nabla_{\nu})[f_1'(R)+f_2'(R)f_3(T)]\nonumber\\
	=2 \pi T_{\mu \nu}+f_2(R)f_3'(T)T_{\mu \nu}+f_2(R)[f_3'(T) p +\frac{1}{2}f_3(T)]g_{\mu \nu}
\end{eqnarray}

Here $T_{\mu \nu}$ is the energy-momentum tensor of matter fields in the space-time, $\makebox = \nabla ^{i} \nabla _{i}$ and the prime denotes differentiation with respect to the argument. We consider $f_1(R)=\lambda _1 R, f_2(R)=\lambda _2 R, f_3(T)=\lambda _3 T$ with $\lambda_1=\lambda_2=\lambda_3=\lambda$.

Considering $(g_{\mu \nu}\makebox-\nabla_{\mu}\nabla_{\nu})=0$, eqn.(4) reduce significantly as,
\begin{eqnarray}
	\lambda [R_{\mu \nu}-\frac{1}{2} R g_{\mu \nu}]=2 \pi T_{\mu \nu}+(\lambda^2 p +\frac{1}{2}\lambda^2 T) R g_{\mu \nu}
\end{eqnarray}
or as,
\begin{eqnarray}
	G_{\mu \nu}-\lambda (p +\frac{1}{2}T) R g_{\mu \nu}=\frac{2 \pi}{\lambda} T_{\mu \nu}
\end{eqnarray}
Equating eqn.(4) with 
\begin{eqnarray}
	G_{\mu \nu}-\Lambda g_{\mu \nu} = - 2\pi T_{\mu \nu},
\end{eqnarray}
we get
\begin{eqnarray}
	\Lambda(R,T)= \lambda (p +\frac{1}{2}T) R
\end{eqnarray}

We have chosen a sufficiently small negative value of $\lambda$ to keep in parity the same sign on the RHS of eqns.(6) and (7). The value of $\lambda$ is considered to remain unchanged throughout.
The above eqn. in the dust case $p=0$, reduce as
\begin{eqnarray}
	\Lambda=\Lambda(R,T)= \frac{1}{2} \lambda R T
\end{eqnarray}
In the case of perfect fluid \cite{b} $T=\rho-2p$ when eqn.(8) reduce as
\begin{eqnarray}
	\Lambda=\frac{1}{2} \rho \lambda R
\end{eqnarray}
In the case of dust universe \cite{a} $T=\rho$ when eqn.(9) reduce as
\begin{eqnarray}
	\Lambda=\frac{1}{2} \rho \lambda R
\end{eqnarray}

In fact we observe that the cosmological constant, if treated as a function of cosmic time 't' remains invariant both for perfect fluid as well as in dust case.

We may think that the expansion of the universe is accelerating due to the positive sign of cosmological constant. Also if one solves the Einstein’s general theory of relativity equation with a positive cosmological constant, one obtains a solution for spacetime with positive Gaussian curvature. So, we may consider to be in a de-Sitter universe. Here for a positive scalar curvature, $R>0$ implies $\rho >0$ for a positive cosmological constant. It may be taken as $\Lambda= 10^{-123}$ acording to present observational data.

\section{Field equations and solutions of the generalized metric in (2+1)-Dimensional Spacetime:}

~~~In $(2 + 1)$ dimensional gravity, we consider the metric,
\begin{equation}
	ds^2 = dt^2 - e^{2 A(t)} dx^2 - e^{2 B(t)} dy^2 ,
\end{equation}

where A(t) and B(t) are functions of t alone.

The energy momentum tensor for perfect fluids is given by,
\begin{eqnarray}
	T_{\mu \nu}= (\rho+p)u_{\mu} u_{\nu}-pg_{\mu \nu}
\end{eqnarray}
where the velocity vector $u^{\mu}=(0,0,1)$ satisfying $u^{\mu} u_{\nu}=1$ and $u^{\mu} \nabla_{\nu} u_{\mu}=0$ , $\rho$ and p are defined as the energy density and pressure of the fluid respectively.

Now the cosmological eqn.(6) for the energy momentum tensor defined in eqn.(13) and the metric in eqn.(12), give rise to the field equations as,
\begin{eqnarray}
	\dot{A} \dot{B}= - \frac{2 \pi}{\lambda} \rho - \Lambda \nonumber\\
	\dot{B}^2 +\ddot{B} = \frac{2 \pi}{\lambda} p - \Lambda \nonumber\\
	\dot{A}^2 +\ddot{A} = \frac{2 \pi}{\lambda} p - \Lambda \nonumber\\
	R=-2(\dot{A}^2+\ddot{A}+\dot{B}^2+\ddot{B}+\dot{A}\dot{B})
\end{eqnarray}
Here a `.' denotes differentiation with respect to the cosmic time t and R is the Ricci scalar.\\

We consider some cases for generating new solutions:\\

{\bf{Case (i)}(a)}: Let $B(t) = k_1 t^n+ \frac{k_2}{t^n}$,\\

where $k_1$ and $k_2$ are positive constants and $n$ is a real number.
The following relations then follow from eqn.(13)
\begin{eqnarray}
	A(t) = k_1 t^n+ \frac{k_2}{t^n}~~~~~~~~\nonumber\\
	\ddot{B} =\ddot{A}= k_1 n (n-1)t^{n-2}+ k_2 n (n+1) t^{-n-2}
\end{eqnarray}
We derive the pressure and density as,
\begin{eqnarray}
	p(t) = \frac{\lambda}{2 \pi}[\Lambda+k_1 n (n-1)t^{n-2}+ k_2 n (n+1)
	\times t^{-n-2}
	+(k_1 n t^{n-1}-k_2 n t^{-n-1})^2]\nonumber\\
	\rho(t) =-\frac{\lambda}{2 \pi}[\Lambda+(k_1 n t^{n-1}-k_2 n t^{-n-1})^2]
\end{eqnarray}
The Ricci scalar as a function of t from eqn.(13) is given as,
\begin{eqnarray}
	R(t)=\frac{4\pi(\rho-2p)}{\lambda (1-3\rho \lambda)}
\end{eqnarray}

\begin{figure}[htbp]
	\centering
	\includegraphics[scale=0.85]{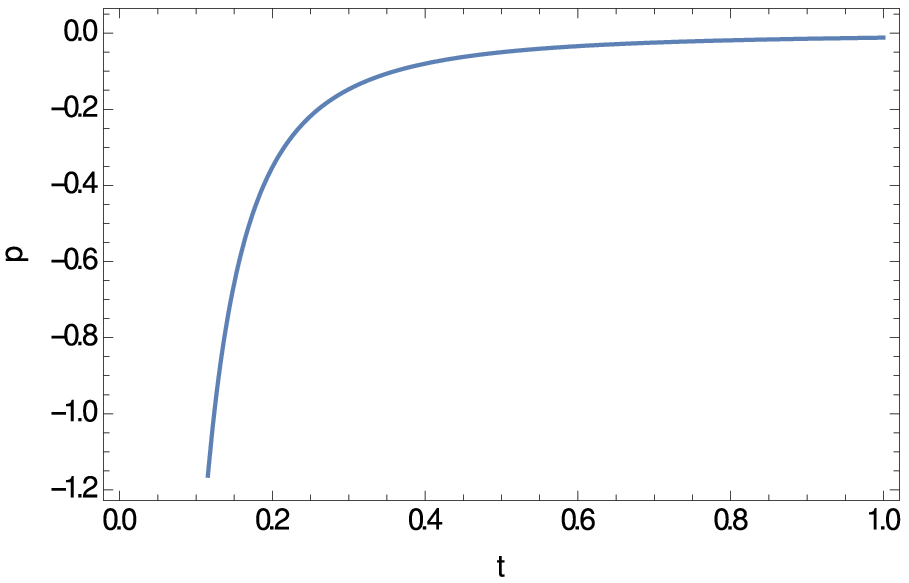}
	\caption{The pressure is plotted against $t$ for the values of constants, $\lambda=-0.1$,$k_1=-1$,$k_2=1$,$n=0.25$,$\Lambda=10^{-123}($taking the cosmological constant as having constant value$)$. }
\end{figure}

\begin{figure}[htbp]
	\centering
	\includegraphics[scale=0.85]{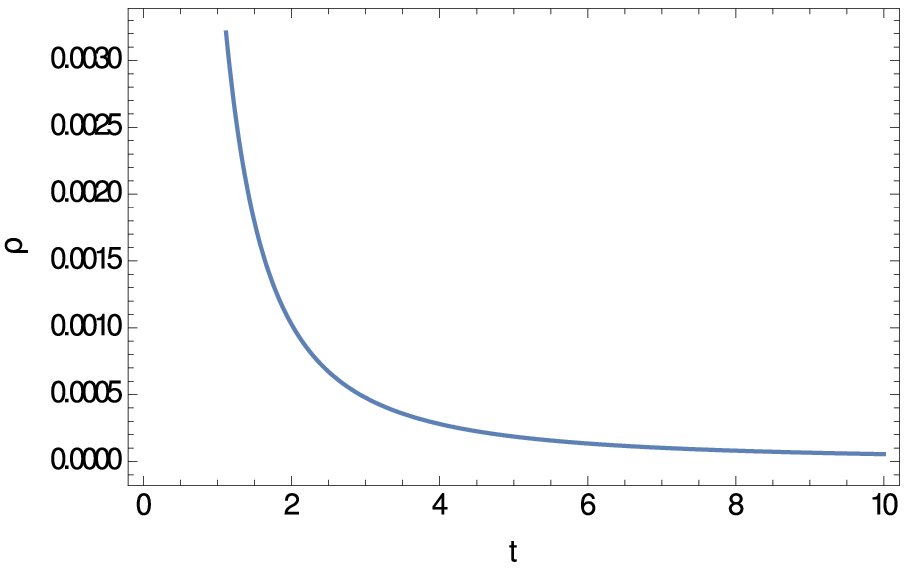}
	\caption{The density is plotted against $t$ for the values of constants, $\lambda=-0.1$,$k_1=-1$,$k_2=1$,$n=0.25$,$\Lambda=10^{-123}($taking the cosmological constant as having constant value$)$.}
\end{figure}

\begin{figure}[htbp]
	\centering
	\includegraphics[scale=0.85]{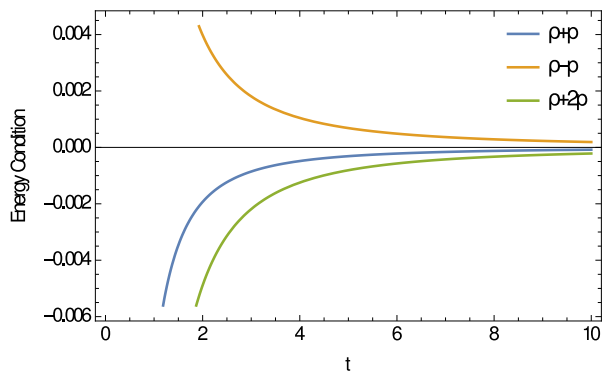}
	\caption{The energy conditions are shown against $t$ for the values of constants, $\lambda=-0.1$,$k_1=-1$,$k_2=1$,$n=0.25$,$\Lambda=10^{-123}($taking the cosmological constant as having constant value$)$.}
\end{figure}

We observe from the Figures.(1-2) that the pressure is negative throughout and $p \to 0$ as $t \to \infty$ whereas the energy density is a positive decreasing function and $\rho \to 0$ as $t \to \infty$. The cosmological constant is taken as positive and $\Lambda=10^{-123}$. As it has very meagre value, the figures does not affect any change if it is considered as having negative value as $\Lambda=-10^{-123}$ or even when $\Lambda=0$.\\

{\bf{(i)}(b)}:
Now we thrive to search for the energy conditions which are given as,\\

(i) Null energy condition (N.E.C) : $\rho+p \geq 0$\\
(ii) Weak energy condition (W.E.C) : $\rho+p \geq 0$, $\rho \geq 0$\\
(iii) Dominant energy condition (D.E.C) : $\rho \geq |p|$\\
(iv) Strong energy condition (S.E.C): $\rho+p \geq 0$, $\rho+2p \geq 0$.

The plot of energy conditions in Fig.(3) imply that only the WEC and DEC are partially satisfied in our model.\\

{\bf{(i)(c)}}: If however, the cosmological constant is treated as a function of $t$, i.e $\Lambda=\Lambda(t)$ then we find it from eqns.(9) and (16) as,
\begin{eqnarray}
	\Lambda(t)= -[\lambda^2 n^3t^{-2n-2} (k_1 t^{2n}-k_2)^2 (-2k_1 t^{3n}
	2k_1 n t^{3n}+ 3k_1^2 n t^{4n}+2 k_2 t^n+2k_2 n t^n\nonumber\\
	-6k_1 k_2 n t^{2n}+3k_2^2 n)\times[-2 \pi t^{2n+2}
	-2 \lambda^2 k_1 n t^{3n}+2k_1 \lambda^2 n^2 t^{3n}+3k_1^2 \lambda^2 n^2 t^{4n}
	+2k_2 \lambda^2 n t^n\nonumber\\+ 2k_2 \lambda^2 n^2 t^n-6k_1 k_2 \lambda^2 n^2 t^{2n}
	+3k_2^2 \lambda^2 n^2]^{-1}
\end{eqnarray}
It is shown graphically below,
\begin{figure}[htbp]
	\centering
	\includegraphics[scale=0.85]{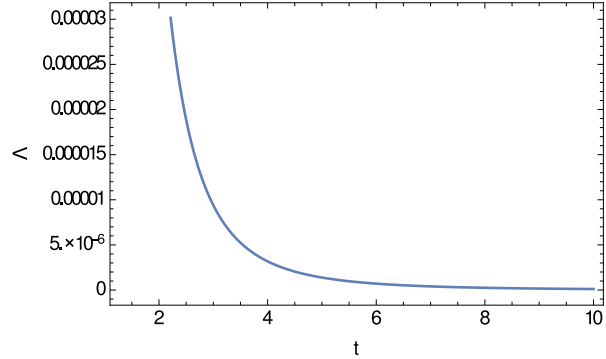}
	\caption{The cosmological constant is shown against $t$ for the values of constants, $\lambda=-0.1$,$k_1=-1$,$k_2=1$,$n=0.25$.}
\end{figure}
We observe that the cosmological constant is positive and decreasing function. However $\Lambda \to 0$ as $t \to \infty$. We observe the disparity between the expected and observed values of $\Lambda$. This is the cosmological constant problem which requires further investigation to explain why it constant assumes such a meagre value today. The corresponding pressure and density terms are,

\begin{eqnarray}
	p(t) = \frac{\lambda}{2 \pi}[k_1 n (n-1)t^{n-2}+ k_2 n (n+1)
	\times t^{-n-2}
	+(k_1 n t^{n-1}-k_2 n t^{-n-1})^2\nonumber\\-\lbrace \lambda^2 n^3t^{-2n-2} (k_1 t^{2n}-k_2)^2 (-2k_1 t^{3n}
	2k_1 n t^{3n}+ 3k_1^2 n t^{4n}+2 k_2 t^n+2k_2 n t^n\nonumber\\
	-6k_1 k_2 n t^{2n}+3k_2^2 n) \times(-2 \pi t^{2n+2}
	-2 \lambda^2 k_1 n t^{3n}+2k_1 \lambda^2 n^2 t^{3n}+3k_1^2 \lambda^2 n^2 t^{4n}
	+2k_2 \lambda^2 n t^n\nonumber\\+ 2k_2 \lambda^2 n^2 t^n-6k_1 k_2 \lambda^2 n^2 t^{2n}
	+3k_2^2 \lambda^2 n^2)^{-1}\rbrace]\nonumber\\
	\rho(t) =-\frac{\lambda}{2 \pi}[(k_1 n t^{n-1}-k_2 n t^{-n-1})^2-\lbrace \lambda^2 n^3t^{-2n-2} (k_1 t^{2n}-k_2)^2 (-2k_1 t^{3n}\nonumber\\
	2k_1 n t^{3n}+ 3k_1^2 n t^{4n}+2 k_2 t^n+2k_2 n t^n
	-6k_1 k_2 n t^{2n}+3k_2^2 n) \times(-2 \pi t^{2n+2}
	-2 \lambda^2 k_1 n t^{3n}\nonumber\\+2k_1 \lambda^2 n^2 t^{3n}+3k_1^2 \lambda^2 n^2 t^{4n}
	+2k_2 \lambda^2 n t^n+ 2k_2 \lambda^2 n^2 t^n-6k_1 k_2 \lambda^2 n^2 t^{2n}
	+3k_2^2 \lambda^2 n^2)^{-1}\rbrace]
\end{eqnarray}
For $\Lambda=\Lambda(t)$ the corresponding pressure, density and energy conditions are shown as below,
\begin{figure}[htbp]
	\centering
	\includegraphics[scale=0.85]{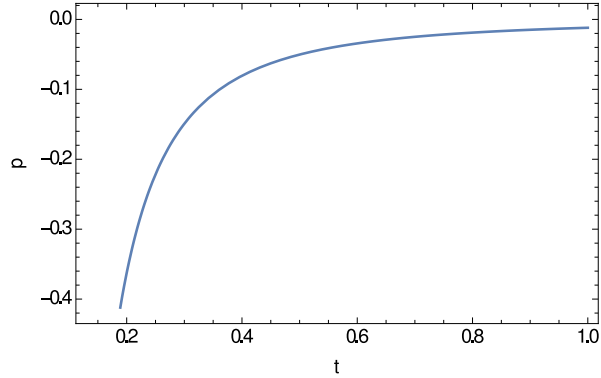}
	\caption{The pressure is plotted against $t$ for the values of constants, $\lambda=-0.1$,$k_1=-1$,$k_2=1$,$n=0.25$}
\end{figure}

\begin{figure}[htbp]
	\centering
	\includegraphics[scale=0.85]{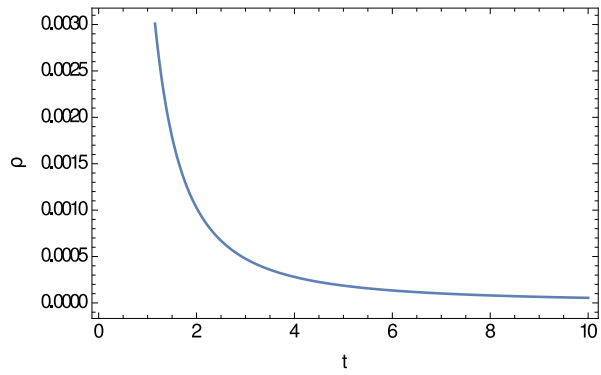}
	\caption{The density is plotted against $t$ for the values of constants, $\lambda=-0.1$,$k_1=-1$,$k_2=1$,$n=0.25$}
\end{figure}
\begin{figure}[htbp]
	\centering
	\includegraphics[scale=0.85]{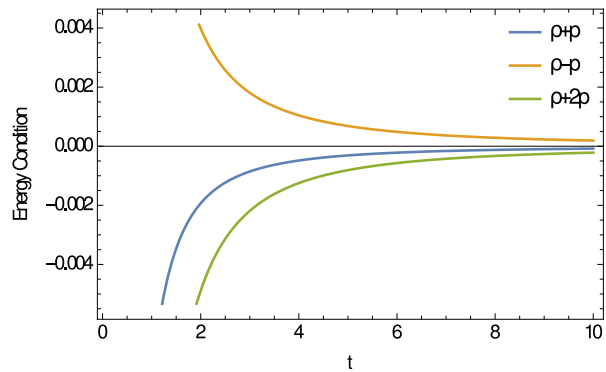}
	\caption{The energy conditions are shown against $t$ for the values of constants, $\lambda=-0.1$,$k_1=-1$,$k_2=1$,$n=0.25$}
\end{figure}
The plot of energy conditions in Fig.(7) imply that only the WEC and DEC are partially satisfied in our model.\\

{\bf{(i)(d)}}: Scale factor a(t), Hubble parameter H(t) and deceleration parameter q(t) with $\Lambda=\Lambda(t)$:\\

The average scale factor a(t), Hubble parameter H(t) and deceleration parameter q(t) using eqn.(14) are derived as \cite{e}.
\begin{eqnarray}
	a(t)= e^{\frac{A(t)+B(t)}{2}}\nonumber\\
	H(t)= \frac{\dot{a}}{a}= \dot{B}=k_1 n t^{n-1}-k_2 n t^{-n-1}\nonumber\\
	q(t)=-(1+\frac{\dot{H}}{H^2})
	=-\lbrace 1+\frac{k_1 n (n-1)t^{n-2}+ k_2 n (n+1) t^{-n-2}}{(k_1 n t^{n-1}-k_2 n t^{-n-1})^2} \rbrace
\end{eqnarray}

\begin{figure}[htbp]
	\centering
	\includegraphics[scale=0.85]{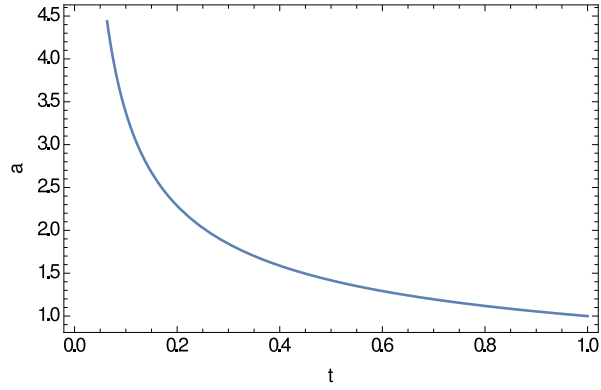}
	\caption{The scale factor is shown against $t$ for the values of constants, $\lambda=-0.1$,$k_1=-1$,$k_2=1$,$n=0.25$}
\end{figure}
\begin{figure}[htbp]
	\centering
	\includegraphics[scale=0.85]{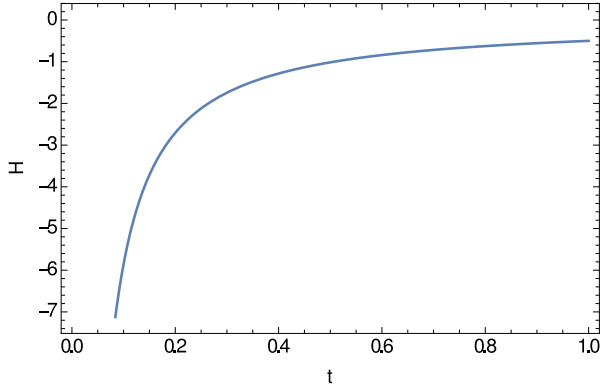}
	\caption{The Hubble parameter is shown against $t$ for the values of constants, $\lambda=-0.1$,$k_1=-1$,$k_2=1$,$n=0.25$}
\end{figure}
\begin{figure}[htbp]
	\centering
	\includegraphics[scale=0.85]{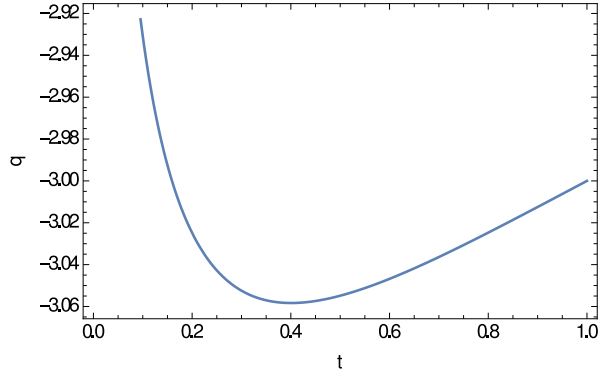}
	\caption{The deceleration parameter is shown against $t$ for the values of constants, $\lambda=-0.1$,$k_1=-1$,$k_2=1$,$n=0.25$}
\end{figure}

We observe that the deceleration parameter first decreases and then increases.\\

\section{Final Remarks}

~~~We have thus developed a feasible cosmological model under $f(R,T)$ gravity. Various physical aspects of the model are elucidated. Present observational results  suggest that our universe can have a non-zero cosmological constant as it is accelerating. Our study too, supports a non-zero cosmological constant. Further study with
$p= \omega \rho$,
where $-1 < \omega < 1$ is in progress.
The cosmological constant can be interpreted as arising from a form of energy which has negative pressure, equal in magnitude to its (positive) energy density where $\omega=-1$.
Such form of energy- a generalization of the notion of a cosmological constant is known as dark energy. On the other hand if $\omega=-\frac{1}{3}$, we get a quintessence field which is a hypothetical form of dark energy, more precisely a scalar field and postulated as an explanation of the observation of an accelerating rate of expansion of the universe.

\section{Acknowledgments}
~~~SI is thankful to S Datta, Md. A Shaikh and P. Tarafdar for providing some useful insights in the paper.


\section*{References}
\begin{thebibliography}{9}

\bibitem{i} Romero C. and Dahia F., Int. J. Theor. Phys., {\bf 33}, 2019 (1994).
\bibitem{m} Cataldo M. and Cruz N., Phys. Rev. D, {\bf 73}, 104026 (2006).	
\bibitem{n} Barrow J. D. et al., Class. Quantum Grav. {\bf 23} 5291 (2006).
\bibitem{o} Rahaman F. et al., Eur. Phys. J. C, {\bf 74}, 2845 (2014).
\bibitem{j} Shamir M. F., Astrophys. Space Sci., {\bf 330}, 183-189 (2010).
\bibitem{p} Bengochea G. R. and Ferraro R., Phys. Rev. D, {\bf 79}, 124019 (2009).
\bibitem{l} Sharif M. and Zubair M., J. Phys. Soc. Jpn. {\bf 82}, 014002 (2013)
\bibitem{g} Perlmutter, S., et al., Nature, {\bf 391} 51 (1998).
\bibitem{h}	Reiss, A.G, et al., Astron. J. {\bf 116} 1009 (1998).
\bibitem{f} Capozziello S. et al., Physics Letters B, {\bf 730}, 280-283 (2014).
\bibitem{c} Pawar D. D. et al., Aryabhatta Journal of Mathematics and Informatics, {\bf 7}, 17 (2015).
\bibitem{d} Houndjo, M.J.S. et al., Can. J. Phys., {\bf 91}, 548 (2013).
\bibitem{k} Houndjo, M.J.S., Int. J. Mod. Phys. D, {\bf 21}, 1250003 (2012).
\bibitem{b} Ahmed N. et al.,  NRIAG Journal of Astronomy and Geophysics, {\bf 5}, 35-47 (2016).
\bibitem{a} Harko T., Phys. Rev. D, {\bf 84}, 024020 (2011).
\bibitem{e} Yadav A. K. et al., Int. J. Theor. Phys., {\bf 50}, 871–881 (2011).  

\end{thebibliography}
\end{document}